\documentclass[11pt,a4paper]{article}

\begin{document}

\title{From Kepler's Laws to Newtonian Motion and the Direction Angle of Hamilton's Hodograph}
\author{Dr. Klaus Huber \\ 
	  Huber Consult \\
           Sesenheimer Str. 21 \\
          10627 Berlin \\ 
          Germany         \\
 email: email@klaus-huber.net}
\date{ }
\maketitle
\begin{abstract}
In this contribution it is shown that the path from Kepler's results to Newtonian motion can be remarkably short and simple.
Following this path we also give a straight forward computation of the direction angle of Hamilton's Hodograph. Then we show
how the speed as function of the direction angle can be expressed and inverted elegantly using elliptic functions.

\vspace{1mm}

\end{abstract}

\section{Introduction}

Usually Newtonian motion of planets under an inverse square law force is established using Kepler's second law.
Asuming at the outset that the inverse square law rules elliptic motion results (or hyperbolic and parabolic motion).

Here we use Kepler's first and second law to find the inverse square law. This procedure also gives the opportunity to
find a simple formula for the direction angle of Hamilton's Hodograph.

\section{Kepler's laws}
\label{sec:Keplerslaws}

We start ab ovo with Kepler's three laws:

\begin{itemize}
\item[i.]   A planet moves on an ellipse around the sun with the sun placed in a focus of the ellipse.
\item[ii.]  The straight line from the sun to a planet sweeps out equal areas in equal times.
\item[iii.] The time period $T$ for a revolution and the major semiaxis $a$ of the ellipse behave as $T^2 = \mbox{const} \cdot a^3$.
\end{itemize}

%\newpage

\section{The shortest path from Kepler's laws to Newtonian motion}

All motion happens in a plane. Without loss of generality we consider an ellipse with reference to perihel (i.e. the position where the distance
of the planet is closest to the sun). The radius is then given by
\[
   r = \frac{a (1-e^2)}{1 + e \cos \theta} \mbox{ . }
\]
The above parameters are: the major semiaxis $a$, the excentricity of the ellipse $e$, and the angle $\theta$ called true anomaly in astronomy.
It is convenient to use complex numbers with $\imath=\sqrt{-1}$. The position in the complex plane is given by
\[
   Z = \frac{a (1-e^2)}{1 + e \cos \theta} \, \exp(\imath \theta) \mbox{ . }
\]
The derivative with respect to time $t$, denoted by a dot, then immediately gives the velocity 
\[
   \dot Z = \frac{a (1-e^2) \dot \theta}{(1 + e \cos \theta)^2} \left\{ \imath e + \imath  \exp(\imath \theta) \right\} \mbox{ . }
\]
From Kepler's second law we find
$\frac{1}{2} r^2 \dot \theta = \frac{\pi a b}{T} $ where $\pi a b$ is the area of an ellipse with major- and minor semiaxes $a$ and $b$ and $T$ is the time
for a complete revolution. As $b=a \sqrt{1-e^2}$ we get
$\dot \theta = \frac{2 \pi}{T (1-e^2)^\frac{3}{2}} (1+e \cos \theta)^2$ and
\begin{equation}
\label{eq:Zdot}
  \dot Z = \frac{2 \pi a}{T} \frac{1}{\sqrt{1-e^2}} \left\{ \imath e + \imath  \exp(\imath \theta) \right\} \mbox{ . }
\end{equation}
We easily see that $\dot Z$ decribes a circle in the complex plane.
Taking the second derivative yields the acceleration
\[
  \ddot Z = -\frac{4 \pi^2 a^3}{T^2} \frac{(1+ e \cos \theta)^2}{a^2 (1-e^2)^2} \, \exp(\imath \theta)  =-\frac{4 \pi^2 a^3}{T^2} \frac{1}{r^2} \, \exp(\imath \theta) \mbox{ , }
\]
which, setting $\frac{4 \pi^2 a^3}{T^2}=G$, leads to the inverse square law force for an object of mass $m$:
\[
    m \ddot Z = -\frac{mG}{r^2} \, \exp(\imath \theta) \mbox{ . }
\]
Note that Kepler's third law $\frac{4 \pi^2 a^3}{T^2}=G$  is obtained automatically.

\section{Hamilton's Hodograph}

\begin{figure}[h]
\caption{\label{fig:Ellipse} Elliptic orbit having $e=0.6$}
\setlength{\unitlength}{3cm}
\hspace{4cm}
\begin{picture}(2,2)
\qbezier(1.8,1)(1.80002,1.10387)(1.75886,1.2026)
\qbezier(1.75886,1.2026)(1.71091,1.31759)(1.61347,1.41078)
\qbezier(1.61347,1.41078)(1.48018,1.53824)(1.28577,1.59777)
\qbezier(1.28577,1.59777)(0.977198,1.69227)(0.675055,1.58483)
\qbezier(0.675055,1.58483)(0.199688,1.41579)(0.2,1)
\qbezier(0.2,1)(0.200312,0.583991)(0.675055,0.415173)
\qbezier(0.675055,0.415173)(0.976941,0.307822)(1.28577,0.402226)
\qbezier(1.28577,0.402226)(1.47982,0.461542)(1.61347,0.589222)
\qbezier(1.61347,0.589222)(1.71076,0.68216)(1.75886,0.797399)
\qbezier(1.75886,0.797399)(1.79998,0.895921)(1.8,1)
\put(1.48,1){\circle*{0.03}}
\qbezier(1.48,1)(1.61435,1.11274)(1.74871,1.22547)
\qbezier(1.74871,1.22547)(1.73879,1.21422)(1.72887,1.20297)
\qbezier(1.74871,1.22547)(1.73591,1.21766)(1.7231,1.20984)
\qbezier(1.48,1)(1.64,1)(1.8,1)
\put(1.56435,1.16274){$\bf{r}$}
\put(1.66435,1.06274){$\theta$}
\qbezier(1.64,1)(1.64,1.00279)(1.6399,1.00558)
\qbezier(1.6399,1.00558)(1.63981,1.00838)(1.63961,1.01116)
\qbezier(1.63961,1.01116)(1.63942,1.01395)(1.63912,1.01672)
\qbezier(1.63912,1.01672)(1.63883,1.0195)(1.63844,1.02227)
\qbezier(1.63844,1.02227)(1.63805,1.02503)(1.63757,1.02778)
\qbezier(1.63757,1.02778)(1.63708,1.03053)(1.6365,1.03327)
\qbezier(1.6365,1.03327)(1.63592,1.036)(1.63525,1.03871)
\qbezier(1.63525,1.03871)(1.63457,1.04142)(1.6338,1.0441)
\qbezier(1.6338,1.0441)(1.63303,1.04679)(1.63217,1.04944)
\qbezier(1.63217,1.04944)(1.63131,1.0521)(1.63035,1.05472)
\qbezier(1.63035,1.05472)(1.6294,1.05735)(1.62835,1.05994)
\qbezier(1.62835,1.05994)(1.6273,1.06253)(1.62617,1.06508)
\qbezier(1.62617,1.06508)(1.62503,1.06763)(1.62381,1.07014)
\qbezier(1.62381,1.07014)(1.62258,1.07265)(1.62127,1.07512)
\qbezier(1.62127,1.07512)(1.61996,1.07758)(1.61856,1.08)
\qbezier(1.61856,1.08)(1.61717,1.08242)(1.61569,1.08479)
\qbezier(1.61569,1.08479)(1.61421,1.08716)(1.61265,1.08947)
\qbezier(1.61265,1.08947)(1.61108,1.09179)(1.60944,1.09405)
\qbezier(1.60944,1.09405)(1.6078,1.09631)(1.60608,1.09851)
\qbezier(1.60608,1.09851)(1.60436,1.10071)(1.60257,1.10285)
\qbezier(1.74871,1.22547)(1.64641,1.44289)(1.5441,1.6603)
\qbezier(1.5441,1.6603)(1.55245,1.64783)(1.56079,1.63537)
\qbezier(1.5441,1.6603)(1.54839,1.64592)(1.55268,1.63155)
\put(1.67641,1.49289){$\bf{v}$}
\end{picture}
\end{figure}
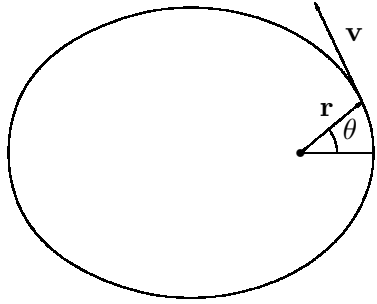

Let $\bf{r}$ be the location vector for the orbit of a planet (see figure~\ref{fig:Ellipse}). Then $\bf{v}=$ $\frac{d}{dt} \bf{r}$ gives the velocity vector, i.e. the vector whose length $v=|\bf{v}|$ gives
the speed. If we consider all vectors $\bf{v}$ on the curve which $\bf{r}$ reaches and if we translate all vectors to the same point (e.g. to the focus for Newtonian motion)
the endpoints of the velocity vectors give rise to a curve which is called the Hodograph. 

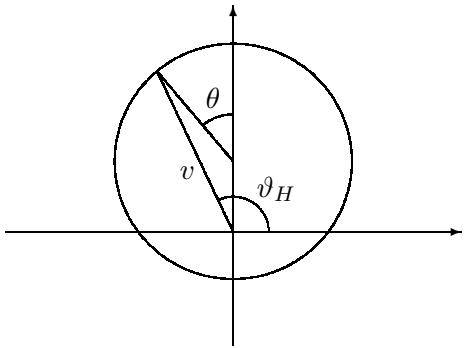
\begin{figure}
\caption{\label{fig:Hodograph} Hodograph $\dot Z= v \cdot \exp(\imath \, \vartheta_H)$ for $e=0.6$}
\setlength{\unitlength}{3cm}
\vspace{1.5cm}
\begin{picture}(2,2)
\put(0,1){\vector(1,0){2}}
\put(1,0.5){\vector(0,1){1.5}}
\qbezier(1.52,1.312)(1.52,1.39436)(1.49455,1.47269)
\qbezier(1.49455,1.47269)(1.4691,1.55102)(1.42069,1.61765)
\qbezier(1.42069,1.61765)(1.37228,1.68428)(1.30565,1.73269)
\qbezier(1.30565,1.73269)(1.23902,1.7811)(1.16069,1.80655)
\qbezier(1.16069,1.80655)(1.08236,1.832)(1,1.832)
\qbezier(1,1.832)(0.91764,1.832)(0.839311,1.80655)
\qbezier(0.839311,1.80655)(0.760982,1.7811)(0.694352,1.73269)
\qbezier(0.694352,1.73269)(0.627721,1.68428)(0.579311,1.61765)
\qbezier(0.579311,1.61765)(0.530901,1.55102)(0.505451,1.47269)
\qbezier(0.505451,1.47269)(0.48,1.39436)(0.48,1.312)
\qbezier(0.48,1.312)(0.48,1.22964)(0.505451,1.15131)
\qbezier(0.505451,1.15131)(0.530901,1.07298)(0.579311,1.00635)
\qbezier(0.579311,1.00635)(0.627721,0.939721)(0.694352,0.891311)
\qbezier(0.694352,0.891311)(0.760982,0.842901)(0.839311,0.817451)
\qbezier(0.839311,0.817451)(0.91764,0.792)(1,0.792)
\qbezier(1,0.792)(1.08236,0.792)(1.16069,0.817451)
\qbezier(1.16069,0.817451)(1.23902,0.842901)(1.30565,0.891311)
\qbezier(1.30565,0.891311)(1.37228,0.939721)(1.42069,1.00635)
\qbezier(1.42069,1.00635)(1.4691,1.07298)(1.49455,1.15131)
\qbezier(1.49455,1.15131)(1.52,1.22964)(1.52,1.312)
\qbezier(1,1.312)(0.832875,1.51117)(0.66575,1.71034)
\qbezier(1,1.52)(0.996369,1.52)(0.992741,1.51987)
\qbezier(0.992741,1.51987)(0.989112,1.51975)(0.985491,1.51949)
\qbezier(0.985491,1.51949)(0.981869,1.51924)(0.978258,1.51886)
\qbezier(0.978258,1.51886)(0.974647,1.51848)(0.971052,1.51798)
\qbezier(0.971052,1.51798)(0.967457,1.51747)(0.963881,1.51684)
\qbezier(0.963881,1.51684)(0.960306,1.51621)(0.956754,1.51545)
\qbezier(0.956754,1.51545)(0.953203,1.5147)(0.94968,1.51382)
\qbezier(0.94968,1.51382)(0.946157,1.51294)(0.942667,1.51194)
\qbezier(0.942667,1.51194)(0.939177,1.51094)(0.935724,1.50982)
\qbezier(0.935724,1.50982)(0.932272,1.5087)(0.92886,1.50746)
\qbezier(0.92886,1.50746)(0.925448,1.50621)(0.922082,1.50485)
\qbezier(0.922082,1.50485)(0.918716,1.50349)(0.915399,1.50202)
\qbezier(0.915399,1.50202)(0.912082,1.50054)(0.908819,1.49895)
\qbezier(0.908819,1.49895)(0.905556,1.49736)(0.90235,1.49565)
\qbezier(0.90235,1.49565)(0.899144,1.49395)(0.896,1.49213)
\qbezier(0.896,1.49213)(0.892856,1.49032)(0.889777,1.48839)
\qbezier(0.889777,1.48839)(0.886698,1.48647)(0.883688,1.48444)
\qbezier(0.883688,1.48444)(0.880678,1.48241)(0.877741,1.48028)
\qbezier(0.877741,1.48028)(0.874803,1.47814)(0.871942,1.47591)
\qbezier(0.871942,1.47591)(0.869081,1.47367)(0.8663,1.47134)
\put(0.8804,1.546){$\theta$}
\qbezier(1,1)(0.832875,1.35517)(0.66575,1.71034)
\qbezier(1.156,1)(1.156,1.00785)(1.15521,1.01566)
\qbezier(1.15521,1.01566)(1.15442,1.02346)(1.15286,1.03115)
\qbezier(1.15286,1.03115)(1.15129,1.03884)(1.14896,1.04634)
\qbezier(1.14896,1.04634)(1.14663,1.05383)(1.14356,1.06105)
\qbezier(1.14356,1.06105)(1.14048,1.06828)(1.1367,1.07515)
\qbezier(1.1367,1.07515)(1.13292,1.08203)(1.12847,1.08849)
\qbezier(1.12847,1.08849)(1.12402,1.09496)(1.11894,1.10094)
\qbezier(1.11894,1.10094)(1.11386,1.10692)(1.10821,1.11237)
\qbezier(1.10821,1.11237)(1.10256,1.11781)(1.09639,1.12266)
\qbezier(1.09639,1.12266)(1.09022,1.12751)(1.08359,1.13171)
\qbezier(1.08359,1.13171)(1.07696,1.13592)(1.06995,1.13944)
\qbezier(1.06995,1.13944)(1.06293,1.14296)(1.0556,1.14575)
\qbezier(1.0556,1.14575)(1.04827,1.14855)(1.04069,1.1506)
\qbezier(1.04069,1.1506)(1.03312,1.15265)(1.02537,1.15392)
\qbezier(1.02537,1.15392)(1.01763,1.1552)(1.0098,1.15569)
\qbezier(1.0098,1.15569)(1.00196,1.15618)(0.994122,1.15589)
\qbezier(0.994122,1.15589)(0.98628,1.15559)(0.978507,1.15451)
\qbezier(0.978507,1.15451)(0.970733,1.15343)(0.963108,1.15158)
\qbezier(0.963108,1.15158)(0.955483,1.14972)(0.948082,1.14711)
\qbezier(0.948082,1.14711)(0.940682,1.1445)(0.93358,1.14115)
\put(1.104,1.156){$\vartheta_H$}
\put(0.766,1.234){$v$}
\end{picture}

\end{figure}

Hamilton introduced the Hodograph in~\cite{Hamilton1847} and found out
that for Newtonian motion according to an inverse square law force the hodograph is a circle. A look at the equation for $\dot Z$ immediately shows that it
describes the hodograph. The only task left is to express $\dot Z$ using the angle $\vartheta_H$ which goes from the perihel to the vector $\bf{v}$,
the subscript $H$ stands for Hamilton. This is displayed in fig.~\ref{fig:Hodograph}.

\noindent For convenience we use the angle $\vartheta=\vartheta_H-\frac{\pi}{2}$ instead of $\vartheta_H$ and find~\footnote{Here and at subsequent places the reader should have no problems to
resolve any ambiguities when applying inverse functions like the $\arctan$-function.}
\begin{equation}
 \tan \vartheta = \frac{\sin \theta}{e + \cos \theta} \mbox{ . }
\end{equation}
To find $\theta$ as function of  $\vartheta$ we express $\dot Z$ using $\vartheta$ which gives
\begin{equation}
\label{eq:ZdotH}
  \dot Z = \frac{2 \pi a}{T} \frac{1}{\sqrt{1-e^2}}  \left( \sqrt{1-e^2 \sin^2(\vartheta) } + e \cos \vartheta  \right)  \cdot  \imath \, \exp(\imath \vartheta) \mbox{ . }
\end{equation}
Equating with~(\ref{eq:Zdot}) and separating real- and imaginary parts gives
\begin{eqnarray*}
  \sin \theta          & = & (\sqrt{1-e^2 \sin^2(\vartheta) } + e \cos \vartheta ) \sin \vartheta \\
   e + \cos \theta  & = &  (\sqrt{1-e^2 \sin^2(\vartheta) } + e \cos \vartheta ) \cos \vartheta
\end{eqnarray*}
which yields
\begin{equation}
 \tan \theta = \frac{\sin \vartheta \, (\sqrt{1- e^2 \sin^2 \vartheta} + e \cos \vartheta)}{\cos \vartheta \, (\sqrt{1- e^2 \sin^2 \vartheta} + e \cos \vartheta ) -e } \mbox{ . }
\end{equation}
Clearly the speed is given by
\begin{equation}
\label{eq:speed}
 v=\frac{2 \pi a}{T} \frac{1}{\sqrt{1-e^2}}  \left( \sqrt{1-e^2 \sin^2(\vartheta) } + e \cos \vartheta  \right)  \mbox{ . }
\end{equation}

\section{Computing $\vartheta$ from $v$}

The speed $v$ as function of $\vartheta$ is given in~(\ref{eq:speed}). We now invert this function, i.e. we solve for $\vartheta$.
This can be done directly using elliptic functions~\footnote{Some basic knowledge of Elliptic functions is assumed, see e.g. the book
of Whittacker and Watson~\cite{Whittaker:Watson}. For the definition of the functions sn, cn, and dn see appendix~\ref{app:A}.}. 
We set $\vartheta = \rm{am}(u,k)$, where $\rm{am}(u,k)$ is Jacobi's amplitude function, which has as input $u$ and the modulus $k$ with $k=e$ and get
\[
 v = \frac{2 \pi a}{T} \frac{1}{\sqrt{1-e^2}} \, \left( \rm{dn}(u,e) + e \cdot \rm{cn}(u,e)  \right) \mbox{ . }
\]
In appendix~\ref{app:A} it is shown that this can be expressed using the $\rm{dn}$-function alone to give
\begin{equation}
\label{eq:speed-dn}
 v = \frac{2 \pi a}{T} \sqrt{\frac{1+e}{1-e}} \cdot \rm{dn} \left( \frac{1+e}{2} u, \frac{ 2 \sqrt{e}}{1+e}  \right)  \mbox{ . } %\hspace{2mm} \mbox{where $\vartheta=\rm{am}(u,e)$. }
\end{equation}
%Noting that the elliptic integral $F(\vartheta,e)=\int_0^\vartheta \frac{d \varphi}{\sqrt{1-e^2 \sin^2 \varphi}}$ is the inverse function of $\rm{am}(u,e)$ we 
Hence we immediately can express $\vartheta$ as function of $v$.
\begin{equation}
 \vartheta = \rm{am}\left( \frac{\rm{dn}^{-1}(\frac{v}{2 \pi a /T} \sqrt{\frac{1-e}{1+e}} , \frac{ 2 \sqrt{e}}{1+e}   )}{(1+e)/2}, e \right) \mbox{ . }
\end{equation}
Note %also 
that all involved functions and inverse functions can be easily and efficiently computed
(in addition they are included in all relevant computer algebra programs).

\section{The angle between $\bf{r}$ and $\bf{v}$}

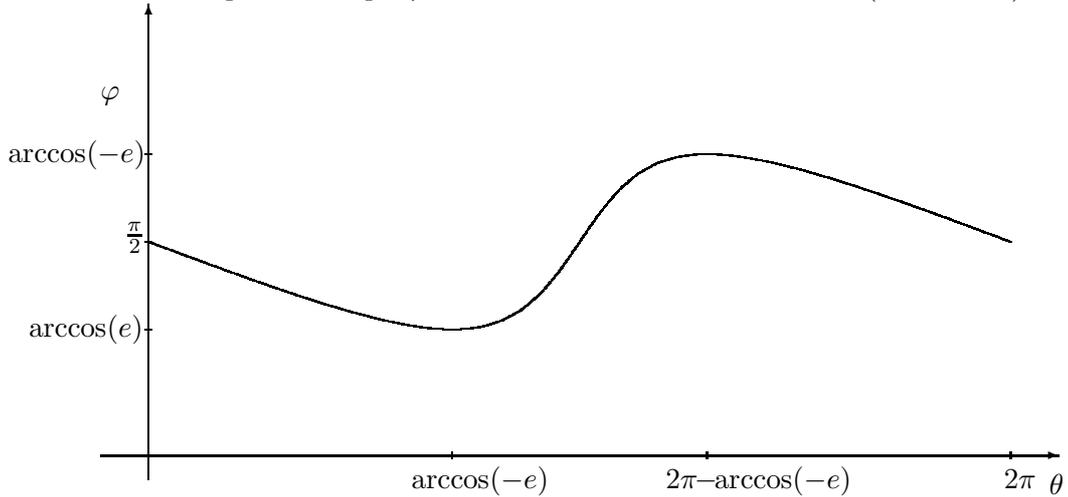
\begin{figure}[h]
\caption{\label{fig:Grafik-phi} Angle $\varphi$ between $\bf{r}$ and $\bf{v}$ as function of $\theta$ (for $e=0.6$)}
\setlength{\unitlength}{0.9mm}
\begin{picture}(140,70)(-9,0)
\put(0,3.5){\vector(1,0){140}}
\put(7,0){\vector(0,1){70}}
\put(132,-1.3){$2\pi$}
\put(133,3){\line(0,1){1}}
\put(51.4045,3){\line(0,1){1}}
\put(45.4045,-1.3){$\arccos(-e)$}
\put(88.5955,3){\line(0,1){1}}
\put(82.5955,-1.3){$2 \pi \! \! - \! \! \arccos(-e)$}
\put(6.5 , 22.0956){\line(1,0){1}}
\put(-10.5 , 20.9956){$\arccos(e)$}
\put(6.5 , 47.9045){\line(1,0){1}}
\put(-13.5 , 46.8045){$\arccos(-e)$}
\put(3.5 ,34.9){$\frac{\pi}{2}$}
\put(6.5 ,35){\line(1,0){1}}
\put(138.6,-2.2){$\theta$}
\put(0,56){$\varphi$}
\qbezier(7,35)(8.43182,34.4631)(9.86364,33.9267)
\qbezier(9.86364,33.9267)(11.2955,33.3911)(12.7273,32.8569)
\qbezier(12.7273,32.8569)(14.1591,32.3244)(15.5909,31.7941)
\qbezier(15.5909,31.7941)(17.0227,31.2665)(18.4545,30.7422)
\qbezier(18.4545,30.7422)(19.8864,30.2216)(21.3182,29.7054)
\qbezier(21.3182,29.7054)(22.75,29.1942)(24.1818,28.6888)
\qbezier(24.1818,28.6888)(25.6136,28.1897)(27.0455,27.698)
\qbezier(27.0455,27.698)(28.4773,27.2144)(29.9091,26.7401)
\qbezier(29.9091,26.7401)(31.3409,26.2761)(32.7727,25.8237)
\qbezier(32.7727,25.8237)(34.2045,25.3844)(35.6364,24.9598)
\qbezier(35.6364,24.9598)(37.0682,24.5519)(38.5,24.1627)
\qbezier(38.5,24.1627)(39.9318,23.7947)(41.3636,23.451)
\qbezier(41.3636,23.451)(42.7955,23.1347)(44.2273,22.8499)
\qbezier(44.2273,22.8499)(45.6591,22.6013)(47.0909,22.3943)
\qbezier(47.0909,22.3943)(48.5227,22.2355)(49.9545,22.1328)
\qbezier(49.9545,22.1328)(51.3864,22.0955)(52.8182,22.135)
\qbezier(52.8182,22.135)(54.25,22.2646)(55.6818,22.5006)
\qbezier(55.6818,22.5006)(57.1136,22.8621)(58.5455,23.3714)
\qbezier(58.5455,23.3714)(59.9773,24.0536)(61.4091,24.9356)
\qbezier(61.4091,24.9356)(62.8409,26.0435)(64.2727,27.3973)
\qbezier(64.2727,27.3973)(65.7045,29.0033)(67.1364,30.844)
\qbezier(67.1364,30.844)(68.5682,32.8703)(70,35)
\qbezier(70,35)(71.4318,37.1297)(72.8636,39.156)
\qbezier(72.8636,39.156)(74.2955,40.9967)(75.7273,42.6027)
\qbezier(75.7273,42.6027)(77.1591,43.9565)(78.5909,45.0644)
\qbezier(78.5909,45.0644)(80.0227,45.9464)(81.4545,46.6286)
\qbezier(81.4545,46.6286)(82.8864,47.1379)(84.3182,47.4994)
\qbezier(84.3182,47.4994)(85.75,47.7354)(87.1818,47.865)
\qbezier(87.1818,47.865)(88.6136,47.9045)(90.0455,47.8672)
\qbezier(90.0455,47.8672)(91.4773,47.7645)(92.9091,47.6057)
\qbezier(92.9091,47.6057)(94.3409,47.3987)(95.7727,47.1501)
\qbezier(95.7727,47.1501)(97.2045,46.8653)(98.6364,46.549)
\qbezier(98.6364,46.549)(100.068,46.2053)(101.5,45.8373)
\qbezier(101.5,45.8373)(102.932,45.4481)(104.364,45.0402)
\qbezier(104.364,45.0402)(105.795,44.6156)(107.227,44.1763)
\qbezier(107.227,44.1763)(108.659,43.7239)(110.091,43.2599)
\qbezier(110.091,43.2599)(111.523,42.7856)(112.955,42.302)
\qbezier(112.955,42.302)(114.386,41.8103)(115.818,41.3112)
\qbezier(115.818,41.3112)(117.25,40.8058)(118.682,40.2946)
\qbezier(118.682,40.2946)(120.114,39.7784)(121.545,39.2578)
\qbezier(121.545,39.2578)(122.977,38.7335)(124.409,38.2059)
\qbezier(124.409,38.2059)(125.841,37.6756)(127.273,37.1431)
\qbezier(127.273,37.1431)(128.705,36.6089)(130.136,36.0733)
\qbezier(130.136,36.0733)(131.568,35.5369)(133,35)
\end{picture}
\end{figure}

Having determined $\vartheta_H$ (or $\vartheta$) we immediately get the angle $\varphi$ between $\bf{r}$ and $\bf{v}$ as
\begin{equation}
\label{eq:varphi_theta}
\varphi = \vartheta_H - \theta = \frac{\pi}{2} + \vartheta - \theta \hspace{3mm} \Rightarrow \hspace{3mm} \varphi = \frac{\pi}{2} + \arctan \left( \frac{\sin \theta}{e + \cos \theta} \right) - \theta \mbox{ . }
\end{equation}
The angle $\varphi$ is displayed as function of $\theta$ in figure~\ref{fig:Grafik-phi} for $e=0.6$.
Clearly, at perihel (at $\theta=0$) and aphel (at $\theta=\pi$) this angle equals $\frac{\pi}{2}$. For $0< \theta < \pi$ the angle $\varphi$ is smaller than $\pi/2$
and in the interval $\pi < \theta < 2 \pi$ the angle $\varphi$ is greater than $\pi/2$.
It is a simple matter to compute the angle $\theta$ at which $\varphi$ reaches its extrema. We get the minimum of $\varphi$ at $\theta=\arccos(-e)$ and the
maximum at $\theta=2 \pi - \arccos(-e)$.
Thus the smallest $\varphi$ equals $\varphi_{\rm{min}}=\pi-\arccos(-e) = \arccos(e)$ and the highest $\varphi$ equals $\varphi_{\rm{max}}=\arccos(-e)$.
From the analysis of~(\ref{eq:varphi_theta}),  for the interval $0 \leq \theta < 2 \pi$ of figure~\ref{fig:Grafik-phi}, the ambiguity of the $\arctan$-function in equation~(\ref{eq:varphi_theta})
is resolved as follows.
\[
   \varphi= \left\{
                           \begin{array}{ll}
                             \frac{\pi}{2}+ \arctan \left( \frac{\sin \theta}{e + \cos \theta} \right) - \theta & \mbox{ for $0 \leq \theta < \arccos(-e)$} \\
                             \frac{3 \pi}{2}+ \arctan \left( \frac{\sin \theta}{e + \cos \theta} \right) - \theta & \mbox{ for $\arccos(-e) \leq \theta < 2 \pi -\arccos(-e) $} \\
                             \frac{5 \pi}{2}+ \arctan \left( \frac{\sin \theta}{e + \cos \theta} \right) - \theta & \mbox{ for $2 \pi -\arccos(-e) \leq \theta < 2 \pi $.} 
                           \end{array}
                  \right. 
\]

\section{Conclusion}

Starting with Kepler's first and second laws it is essentially a three line derivation to obtain Newton's gravitational law.
In addition, from the computation one essentially gets Hamilton's hodograph as side effect. The computation of the speed
as function of the angle $\vartheta$ is a straight-forward task as well as the dependency of $\vartheta$ from the true anomaly $\theta$ (and its inversion).
Using elliptic functions one can also express elegantly the speed and invert it, i.e. express $\vartheta$ as function of
the speed. The new expression of the speed (eqn.(\ref{eq:speed-dn})) is essentially an application of the 
Gauss transform~\footnote{The Gauss transform is the inverse of the Landen transform.} to the
elliptic function $\rm{dn}(u,k)+k \cdot \rm{cn}(u,k)$. The Gauss transform changes the modulus $k$  of the elliptic function to $2 \sqrt{k}/(1+k)$.

\appendix

\section{Elliptic Functions used and Derivation of an Identity}
\label{app:A}
We use the standard notation of Gudermann (see~\cite{Whittaker:Watson}, p.494):
\begin{eqnarray*}
 \sin \rm{am} (u,k) &=& \rm{sn}(u,k) \\
 \cos \rm{am} (u,k) &=& \rm{cn}(u,k) \\
 \sqrt{1-k^2 \, \rm{sn}^2(u,k)} &=& \rm{dn}(u,k) \mbox{ . }
\end{eqnarray*}

\noindent We now derive a formula which leads to equation~(\ref{eq:speed-dn}).
Using the transformation formulas 8.152 from~\cite{Gradsteyn:Ryzhik}, p.915 (the 7-th row) -- where $k'=\sqrt{1-k^2}$, $k_1=\frac{1-k'}{1+k'}$, and $u_1=(1+k')u$ -- we find
\[
 \rm{dn}(u_1,k_1) + k_1 \, \rm{cn(u_1,k_1)} = \frac{1-(1-k') \rm{sn}^2(u,k)}{dn(u,k)} + \frac{1-k'}{1+k'} \, \frac{1-(1+k') \rm{sn}^2(u,k)}{dn(u,k)} 
\]
which leads to
\[
   \rm{dn}(u_1,k_1) + k_1 \, \rm{cn(u_1,k_1)} = \frac{2}{1+k'} \, \rm{dn}(u,k)
\]
If we set $k_1=e$ it follows that $k'=\frac{1-e}{1+e}$, $k=\frac{2 \sqrt{e}}{1+e}$, and $\frac{2}{1+k'}=1+e$ hence
\[
 \rm{dn}(u_1,e) + e \, \rm{cn(u_1,e)} = (1+e) \rm{dn} \left( \frac{1+e}{2} u_1,\frac{2 \sqrt{e}}{1+e} \right)
\]
To avoid collision with the different use of $u$ above we replace $u_1$ by $z$ and get
\[
 \frac{ \rm{dn}(z,e) + e \, \rm{cn(z,e)}}{\sqrt{1-e^2}}  =   \sqrt{\frac{1+e}{1-e}}   \, \rm{dn} \left( \frac{1+e}{2} z,\frac{2 \sqrt{e}}{1+e} \right) \mbox{ . }
\]

%>>>>>>>>>>>>>>>>>>>>>>>>>>>>>>>>>>>>>>>>>>>>>>>>>>>>>>>>>>>>>>>>>>>>>>>>>>>>>


\begin{thebibliography}{99}

\bibitem{Gradsteyn:Ryzhik}  I.S.Gradsteyn, I.M.Ryzhik, "Tables of Integrals, Series, and Products", Academic Press, Sixth Printing, 1979. 

\bibitem{Hamilton1847} W.R.Hamilton, "New mode of geometrically conceiving, and of expressing in symbolical language, the Newtonian
	law of attraction, and the mathematical problem of determining the orbits and perturbations of bodies which are governed in their motions
	by that law", Proceedings of the Royal Irish Academy, Vol. III, 1847, pp. 344-353.

\bibitem{Whittaker:Watson} E.T.Whittaker, G.N.Watson, "A course of modern analysis", 
	fourth edition, Cambridge, reprinted 1969.



\end{thebibliography}
\end{document}